# Measuring Asset Composability as a Proxy for DeFi Integration

*Research in Progress*

Victor von Wachter[†], Johannes Rude Jensen[†,§], Omri Ross[†,§]

[†]University of Copenhagen, Department of Computer Science
[§]eToroX Labs, Denmark

`victor.vonwachter@di.ku.dk`

**Abstract.** Decentralized financial (DeFi) applications on the Ethereum blockchain are highly interoperable because they share a single state in a deterministic computational environment. Stakeholders can deposit claims on assets, referred to as 'liquidity shares', across applications producing effects equivalent to rehypothecation in traditional financial systems. We seek to understand the degree to which this practice may contribute to financial integration on Ethereum by examining transactions in 'composed' derivatives for the assets DAI, USDC, USDT, ETH and tokenized BTC for the full set of 344.8 million Ethereum transactions computed in 2020. We identify a salient trend for 'composing' assets in multiple sequential generations of derivatives and comment on potential systemic implications for the Ethereum network.

## 1    Introduction

Smart contracts on the Ethereum blockchain share a single state in a deterministic execution environment [1], a feature which introduces a high level of interoperability between decentralized financial (DeFi) applications. This novelty has thus far, resulted in a rich ecosystem of financial applications, primarily lead by peer-to-peer borrowing/lending markets [2][3] and constant function market makers (CFMM) [4][5]. At the time of writing, crypto assets valued in excess of $39 billion is managed by some 75[1] decentralized financial (DeFi) applications on the Ethereum blockchain.

---

[1] defipulse.com, as of 31st Jan 2020

From the consumers' perspective, interoperability between financial applications is a desirable feature, resulting in a vibrant and highly competitive marketplace of increasingly exotic financial products. Yet, if left unsupervised, interoperability between liquidity reserves may lead to dependencies amongst applications, as techniques equivalent to the practice of rehypothecation in the traditional financial system [6] become normalized.

When allocating assets to a CFMM such as Uniswap, Curve or Balancer, liquidity providers receive 'liquidity provider shares' (LP shares) [7] redeemable for a proportional share of the liquidity pool with the unrealized returns of the position. LP shares are typically computed as transferable, fungible tokens which has led to the emergence of new secondary markets in which applications offer liquidity and lending pools for LP shares themselves. Supplying LP shares to these pools results in the issuance of *meta* LP shares. This process is, in some cases, repeated recursively as stakeholders seek to maximize yield or functionality across a diverse set of applications. While LP shares are often treated by market participants as simple IOUs, they do in fact represent a complex pay-out function, as shown in the literature by [7][8]. Further complicating matters, the practice of 'yield farming', i.e. allocating assets across DeFi applications to maximize returns [9], has introduced a competitive environment in which applications seek to attract additional liquidity by rewarding LP shareholders with 'governance tokens' [10].

We approach Ethereum as a financial ecosystem with structural properties comparable to those of a single market [11][12]. For this work, we examine the degree to which a crypto asset can be utilized in a sequence of increasingly complex 'wrapping' operations, guiding our research question:

*Can we measure assets composability as a proxy for financial integration on the Ethereum Blockchain?*

Informed by the process proposed by [13], we measure the degree to which crypto assets in smart contracts may contribute towards effects equivalent to financial integration on the Ethereum blockchain. We approach transaction data on Ethereum with an asset oriented perspective, in contrast to previous studies of financial activity on Ethereum, sorting by addresses [13] or applications [14].

## 2 Method

We measure asset composability by identifying the number of derivatives produced from an initial root asset *I*. We extend work presented in [13] by proposing an algorithm for unwrapping crypto assets. The algorithm builds a tree structure of derivatives from the initial asset $I$ (Figure 1.). We measure the distance δ to the initial asset as $\delta_{A \to I} =$

```
1: repeat
2:      T <- all transactions of initial assets from block #9193266 to #11565018
3:          draw 10,000 random transactions t in T
4:              for each t:
5:                  identify erc20 tokens in transaction
6:                      if token A is wrapped version of initial asset
7:                          if A less than 100 transfers ignore
8:                              else w_i +1 and calculate distance δ_A
9:                      end if
10:             end for
11: until no relevant new wrapped assets
```

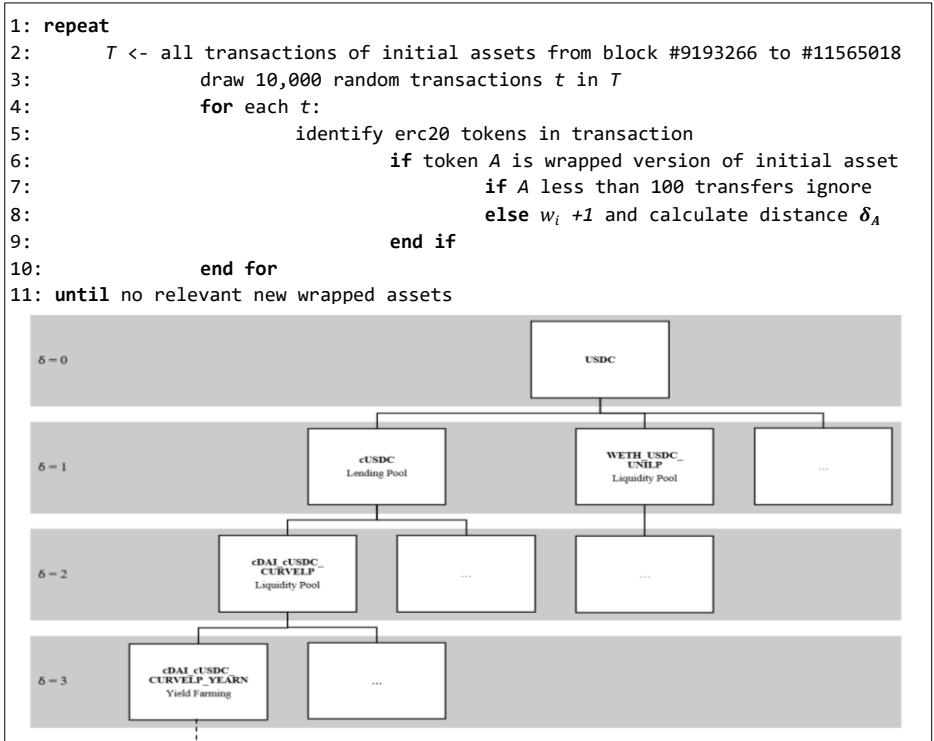

*Figure 1. Method and exemplary asset tree structure*

$\sum_{i=0}^{N} |w_i|$ as a proxy for the degree to which an asset contributes towards integration on Ethereum. That is, the sum of relevant wrapping operations, where $w := (w_1, \dots, w_n)$ is the vector of all adjustments for the composed asset $A$.

In the example (Figure 1.), an asset is allocated to a CFMM liquidity pool, triggering the issuance of the corresponding LP shares. At this point, we consider the initial asset as wrapped *once*, resulting in a distance of 1. Subsequently allocating the LP share to another application would trigger the issuance of another LP share, which amounts to a distance of $\delta_{A \to I} = 2$. We target five popular crypto assets: DAI, USDT, USDC, ETH, and tokenized BTC for the duration of 2020 (Table 1.). Collectively, the selected assets amounted to >70% of the total value administered within DeFi applications[2] at the end of the sample period.

| Initial asset | Transactions on Ethereum | Transactions of composed assets |
|---|---|---|
| DAI | 4,149,654 | 1,033,674 |

---

[2] defipulse.com, as of 31st Jan 2020

| | | |
|---|---|---|
| USDT | 64,956,383 | 687,705 |
| USDC | 7,053,402 | 1,167,163 |
| WETH | 21,187,823 | 919,165 |
| BTC (WBTC, renBTC, sBTC)[3] | 658,035 | 193,394 |

*Table 1. Transactions of assets and composed versions in 2020*

## 3 Results

We find derivatives of the five initial assets among all 344.8 million Ethereum transactions in 2020 (block #9193266 to #11565018). For each initial asset we compare the number of transactions in the 'plain vanilla' version of the asset, against the number of transactions in its derivatives (Figure 2.).

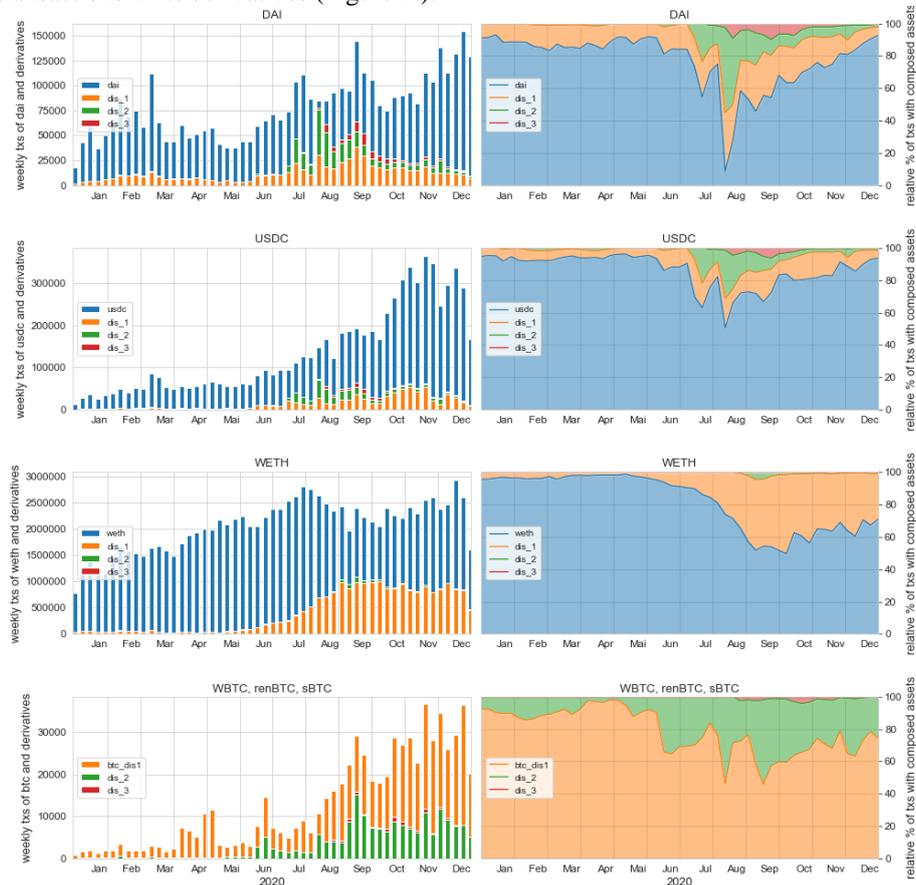

*Figure 2. Financial integration of assets during 2020*

---

[3] Bitcoin (BTC) is a non-native asset on Ethereum, represented by 'wrapped bitcoin' locked on the original blockchain. We compile the three largest representations of Bitcoin on Ethereum into a single category, assigning the category an initial distance of one.

For the first 6 months plain DAI transfers amounted between 82% - 91% (blue) of all DAI asset transfers and composed DAI with $\delta_{A \to I} = 1$ amounted between 9% - 18% (orange) respectively. The data indicates a clear trend towards increasingly complex wrapping operations peaking in the third quarter of 2020, a period colloquially referred to as 'DeFi Summer' due to a high volume of governance tokens issued at the time [10]. The tendency is especially salient in 'DAI', for which to up to 84% of all transactions involved a 'wrapped' derivative of the initial asset.

Curiously, the asset with the largest market cap on Ethereum[4], USDT, appears to be the least popular with an insignificant 687,705 transactions in 'wrapped' derivatives, compared to 64,956,383 transactions in the plain asset.

## 4   Discussion and Conclusions

Computing fractional ownership claims in a deterministic, single state environment introduces a large set of new opportunities for innovation in the financial sector. Because transactions on permissionless blockchains, such as Ethereum, settles atomically, the role for central clearing counterparties in mitigating counterparty risk is largely mitigated for simple transactions. Yet, to date, little is understood about the systemic implications of the design of these applications and how novel concepts like LP shares, may exacerbate the impact of shocks triggered by exploits [15][16].

A quantifiable approach to the study of financial integration on the Ethereum network will facilitate a better understanding how shocks travel through tightly interconnected webs of DeFi applications, which may provide guidance towards promoting resilience and protecting investors against systemic risk. In this work, we present initial indicators by examining the degree to which transactions in 'wrapped' derivatives of an asset, representing increasingly complex payout functions, may offer an indication of the degree of financial integration on the network. We position this contribution within the broader literature on the quantification of 'composability risk' for the DeFi ecosystem, a critical gap raised by [6].

To provide actionable insights for market participants and regulators, this and future studies must expand the scope by considering all relevant factors for the transmission of shocks, including smart-contract design and default risk for the individual DeFi application.

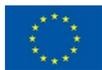 This project has received funding from the European Union's Horizon 2020 research and innovation programme under the Marie Skłodowska-Curie grant agreement No 801199

---

[4] $26.5 billion, as of 31st Jan 2020.